\documentclass[runningheads]{llncs}

\newif\ifAnonymous

\usepackage[utf8]{inputenc}

\usepackage[T1]{fontenc}
\usepackage{xcolor}
\usepackage{listings}
\usepackage{inconsolata}
\usepackage{relsize}
\usepackage{amsmath,amssymb,array}
\usepackage{esz}
\usepackage{mathpartir}
\usepackage{graphicx}
\usepackage[figuresright]{rotating}
\usepackage{pgfplots}
\pgfplotsset{compat=1.18}
\usepgfplotslibrary{groupplots}
\usepackage{xspace}
\usepackage{xurl}
\usepackage{pifont}
\usepackage{accsupp}
\DeclareRobustCommand{\Checkmark}{\BeginAccSupp{ActualText=ok}\ding{51}\EndAccSupp{}}
\DeclareRobustCommand{\XSolidBrush}{\BeginAccSupp{ActualText=failed}\ding{55}\EndAccSupp{}}
\usepackage{tabularx}
\usepackage{enumitem}
\usepackage{booktabs}
\usepackage{microtype}

\usepackage{tcolorbox}
\tcbuselibrary{listings,breakable,skins}

\usepackage[bookmarks,unicode,colorlinks=true,hypertexnames=false,
            linkcolor=blue,citecolor=blue,urlcolor=blue]{hyperref}

\newcommand{\Section}[1]{\section{#1}}
\newcommand{\SubSection}[1]{\subsection{#1}}
\newcommand{\Paragraph}[1]{\subsubsection{#1.}}

\newlist{Description}{description}{1}
\setlist[Description]{leftmargin=1em, labelindent=1em, style=unboxed, font=\bfseries, itemsep=0.5ex, topsep=0.5ex}

\newlist{Itemize}{itemize}{1}
\setlist[Itemize]{label=--, leftmargin=1em, labelindent=0pt, labelsep=0.5em, itemsep=0.5ex, topsep=0.5ex}

\newcommand{\Anon}[2]{\ifAnonymous #2\else #1\fi}

\newcommand{\MVP}[0]{\textsf{MVP}\xspace}

\newcommand{\WP}[0]{WP\xspace}

\newcommand{\sat}{\mathrel{{\scriptstyle\vert}\mkern-1mu{\scriptstyle\sim}}}

\lstdefinestyle{MoveStyle}{
    basicstyle=\ttfamily,
    basewidth=0.5em,
    keywordstyle=\color{black}, 
    commentstyle=\color{gray}\normalfont\itshape,
    escapechar=@, 
    breaklines=true,
    literate={|~}{{$\sat$\,}}1
        {<=}{{$\leq$}}2
        {>=}{{$\geq$}}2
        {==}{{$=$}}2
        {!=}{{$\neq$}}2
        {==>}{{$\Longrightarrow$}}3
        {forall}{{$\forall$}}1
        {exists}{{$\exists$}}1
        {exists<}{{exists<}}7,
}

\lstdefinelanguage{Move}{
    morekeywords={
        abort,
        aborts_if,
        aborts_of,
        acquires,
        address,
        apply,
        as,
        assert,
        assume,
        borrow_global,
        borrow_global_mut,
        break,
        const,
        continue,
        copy,
        copyable,
        define,
        drop,
        else,
        ensures,
        ensures_of,
        exists,
        false,
        forall,
        friend,
        fun,
        global,
        has,
        havoc,
        if,
        in,
        include,
        invariant,
        key,
        let,
        loop,
        match,
        modifies,
        modifies_of,
        module,
        move,
        move_from,
        move_to,
        mut,
        native,
        num,
        old,
        onabort,
        pragma,
        proof,
        public,
        requires,
        requires_of,
        resource,
        result_of,
        return,
        schema,
        script,
        signer,
        spec,
        split,
        store,
        struct,
        true,
        u8,
        u64,
        u128,
        u256,
        update,
        use,
        with,
        where,
        while},
    sensitive=true,
    morecomment=[l]{//},
    morecomment=[s]{/*}{*/},
}

\lstMakeShortInline[language=Move,style=MoveStyle,breaklines=true,breakatwhitespace=true,escapechar=@]|
\lstMakeShortInline[language=Move,style=MoveStyle,breaklines=true,breakatwhitespace=true,escapechar=@]!

\lstnewenvironment{Move}{
    \lstset{
        language=Move,
        style=MoveStyle,
        basicstyle=\relsize{-1}\ttfamily,
        keywordstyle=\color{blue},
    }
}{
}

\lstnewenvironment{MoveBox}{
    \lstset{
        language=Move,
        style=MoveStyle,
        basicstyle=\relsize{-1}\ttfamily,
        keywordstyle=\color{blue},
    }
}{
}

\definecolor{ModelSol}{HTML}{2A78D6}
\definecolor{ModelTerra}{HTML}{EB6834}
\definecolor{CellMuted}{HTML}{898781}
\definecolor{CellGrid}{HTML}{E1E0D9}
\definecolor{CellAxis}{HTML}{C3C2B7}

\definecolor{UserRule}{HTML}{3F8A4E}
\definecolor{UserBack}{HTML}{F2F9F3}
\definecolor{AgentRule}{HTML}{2C6FB5}
\definecolor{AgentBack}{HTML}{F2F7FC}
\definecolor{ToolRule}{HTML}{B07A1E}
\definecolor{ToolBack}{HTML}{FCF8EF}

\tcbset{
    ProvenanceBox/.style={
        enhanced,
        listing only,
        listing options={
            language=Move,
            style=MoveStyle,
            basicstyle=\relsize{-1}\ttfamily,
            keywordstyle=\color{blue},
        },
        breakable,
        sharp corners,
        boxrule=0pt,
        leftrule=1.5pt,
        boxsep=0pt,
        left=6pt, right=3pt, top=3pt, bottom=3pt,
    },
    ProvenanceTag/.style 2 args={
        overlay unbroken and first={%
            \node[anchor=north east, inner xsep=4pt, inner ysep=2pt,
                  font=\scriptsize\sffamily, text=#1] at (frame.north east) {#2};},
    },
}

\newtcblisting{MoveBoxUser}{
    ProvenanceBox,
    colback=UserBack, colframe=UserRule,
    ProvenanceTag={UserRule}{user},
}

\newtcblisting{MoveBoxAgent}{
    ProvenanceBox,
    colback=AgentBack, colframe=AgentRule,
    ProvenanceTag={AgentRule}{agent},
}

\newtcblisting{MoveBoxTool}[1]{
    ProvenanceBox,
    colback=ToolBack, colframe=ToolRule,
    ProvenanceTag={ToolRule}{#1 tool},
}

\lstnewenvironment{MoveBoxNumbered}{
    \lstset{
        language=Move,
        style=MoveStyle,
        basicstyle=\relsize{-1}\ttfamily,
        keywordstyle=\color{blue},
        numbers=left,
        numberstyle=\scriptsize\color{gray},
    }
}{
}

\lstnewenvironment{MoveDiag}{
    \lstset{
        style=MoveStyle,
        basicstyle=\relsize{-2}\ttfamily,
    }
}{
}

\lstnewenvironment{claude}{
    \lstset{
        basicstyle=\relsize{-1}\ttfamily,
        breaklines=true,
        breakautoindent=true,
        breakatwhitespace=true,
        breakindent=1.2em,
        columns=fullflexible,
        keepspaces=true,
        escapechar=@,
        extendedchars=true,
        inputencoding=utf8,
        literate=
            {⎿}{{$\llcorner$}}1
            {└}{{$\llcorner$}}1
            {├}{{$\vdash$}}1
            {─}{{-}}1
            {│}{{$\vert$}}1
            {◻}{{$\square$}}1
            {◼}{{$\blacksquare$}}1
            {☐}{{$\square$}}1
            {☑}{{$\boxtimes$}}1
            {●}{{$\bullet$}}1
            {○}{{$\circ$}}1
            {◯}{{$\bigcirc$}}1
            {⏺}{{$\bullet$}}1
            {✽}{{$\ast$}}1
            {★}{{$\star$}}1
            {☆}{{$\star$}}1
            {…}{{\ldots}}1
            {↗}{{$\nearrow$}}1
            {↘}{{$\searrow$}}1
            {↳}{{$\hookrightarrow$}}1
            {✓}{{\Checkmark}}1
            {✗}{{\XSolidBrush}}1
            {•}{{$\bullet$}}1
            {–}{{--}}1
            {—}{{---}}1
            {‘}{{`}}1
            {’}{{'}}1
            {“}{{``}}1
            {”}{{''}}1,
        moredelim=[l][\color{blue}\bfseries]{>},
    }
}{
}

\newcommand{\CellTaskLabels}{\detokenize{keep_alive},\detokenize{is_order_valid},\detokenize{available_transaction_queue_capacity},\detokenize{can_execute_with_timelock},\detokenize{is_taker_order},\detokenize{refill},\detokenize{request},\detokenize{validate_enough_stake},\detokenize{find_min_stake_required},\detokenize{update_config},\detokenize{match_order_and_get_next_from_bulk_order},\detokenize{reinsert_order_into_bulk_order},\detokenize{new_bulk_order_request_with_sanitization},\detokenize{validate_scheme},\detokenize{upsert_provider_jwks},\detokenize{get_pending_transactions},\detokenize{partition},\detokenize{select},\detokenize{split_signature_bytes},\detokenize{new_tiers},\detokenize{validate_tiers},\detokenize{cancel_at_price_level},\detokenize{discard_price_crossing_levels},\detokenize{validate_not_zero_sizes},\detokenize{validate_price_ordering},\detokenize{range_with_step}}
\newcommand{\CellLoopFree}{13}

\title{Hybrid Mechanical and Agentic\texorpdfstring{\\}{} Specification Inference}
\titlerunning{Hybrid Mechanical and Agentic Specification Inference}

\Anon{%
	\author{Wolfgang Grieskamp \and Teng Zhang \and Vineeth Kashyap}
	\authorrunning{W. Grieskamp et al.}
	\institute{Aptos Labs, Palo Alto, USA\\ \email{wg@aptoslabs.com}}
}{%
	\author{Anonymous Author(s)}
	\institute{Anonymous Institution(s)}
}

\begin{document}

\maketitle

\begin{abstract}
Adoption of formal verification has long been hindered by the cost of manual specification.
To prove high-level properties, one must also write contracts for supporting functions, many of which are mechanical and low-level.
Specification inference automatically derives such contracts from the program.
We present a hybrid of mechanical and agentic inference for Move, a smart contract language, integrated into its verifier, the Move Prover.
A large language model (LLM) agent drives inference and verification through a tool that calls the Move Prover and derives specifications from code by weakest-precondition (\WP) analysis.
We evaluate our approach on 26 targets, 24 of them extracted from public Move functions of the Aptos framework that have no specifications, under a frontier model and a mid-tier model of the same family (Sol~5.6 and Terra~5.6).
With and without \WP, the agent completes over 90\% of its runs, so agentic inference is effective on its own.
Adding \WP\ lowers its cost by 11--21\% overall and by 40--64\% on targets without loops.

\end{abstract}

\Section{Introduction}
\label{sec:intro}

A roadblock to the adoption of program verification is the \emph{cost of manual specification}.
Developers care about high-level properties, e.g., that a call chain aborts only under stated preconditions, an entrypoint satisfies its postconditions, stored data respects an invariant.
But to prove any of them, the supporting functions need their own contracts, which are often boilerplate, as they restate what the code does.
\emph{Specification inference} derives such contracts from the program automatically, so that developers instead spend effort on the properties worth stating manually.
The classical engine for mechanical inference is Dijkstra's weakest-precondition analysis (\WP)~\cite{dijkstra1975wp}: propagated backward through a function, it yields the exact abort condition and the exact postcondition. However, it depends on loop invariants to provide an induction hypothesis~\cite{BL05}, and despite a long history of research~\cite{CousotHalbwachs78}, there is no complete technique for synthesizing them.

This paper describes a novel combination of agentic specification inference and mechanical weakest-precondition analysis.
An LLM agent conducts the inference, supplies loop invariants, and calls \WP and the verifier as tools.
The approach is implemented for the Move smart contract language, which is similar to safe Rust, especially in its reference semantics.
Move has strong verification support through the Move Prover (\MVP)~\cite{DillGPQXZ22,GrieskampEtAl26HO}.
Due to Move's reference semantics, \MVP can eliminate references~\cite{DillGPQXZ22}.
We run classical \WP on the reference-eliminated Move bytecode and restate its result in the predicate calculus of the specification language, over the original references.
\WP is fully modular because of \emph{behavioral predicates}~\cite{GrieskampEtAl26HO}, which let a specification refer to the contract of another function as a black box.

We evaluate our combination on a corpus of 26 inference targets.
Twenty-four of these targets are extracted from public Move code of the Aptos framework.
These have no prior specifications: a model may have seen this code during training, but it cannot recall a specification that did not exist.
We compare three tactics:
\textit{agent-only}, in which the agent works without the \WP tool;
\textit{hybrid-guided}, in which it follows a prescribed workflow that starts with \WP;
and \textit{hybrid-flexible}, in which it decides itself whether and when to call \WP.
We run each tactic under two models of one family, the frontier model Sol~5.6 and the mid-tier model Terra~5.6.
Completeness is checked with withheld mutants.

Agent-only already does well: it completes 99\% of its runs under the frontier model and 93\% under the mid-tier model.
The two hybrid tactics reach about the same completeness at 11--21\% lower cost.
The saving comes from the targets without loops, where \WP derives the complete contract, and thus the hybrids cost 36--60\% of agent-only.
On the loop targets, the agent still has to find the invariants, and the quantified \WP output adds to the context it re-reads on every turn, so the hybrids cost 9--22\% more on average.
The mid-tier model costs 53--59\% as much as the frontier model.
It falls behind mainly on one target, whose contract needs a permutation argument proved with lemmas.
Under hybrid-guided, the mid-tier model completes 97 of 104 runs; the frontier model under agent-only completes 103 but costs more than twice as much.

\Paragraph{Contributions}
Our paper makes the following contributions:
\begin{Itemize}
\item A \WP analysis over reference-eliminated Move bytecode, whose derived conditions are carried back to the references and global resources of the source, with state labels naming the intermediate memories of a function (\S~\ref{sec:wp}).
\item \WP output designed to be read by an agent rather than passed to a solver: loop-invariant evidence from bounded unrolling, and markers for conditions that are expensive for the solver or incomplete because a callee lacks a contract (\S~\ref{sec:diagnostics}).
\item The combination of the \WP analysis with a coding agent, delivered as a plugin of skills, hooks, and tools, in which the agent supplies what \WP cannot (\S\S~\ref{sec:example} and~\ref{sec:skills}).
\item An evaluation on 26 targets, mostly from real-world code, under three tactics and two models, in which an inferred contract counts as complete only if it kills every withheld mutant (\S~\ref{sec:evaluation}).
\end{Itemize}

The Move Prover was developed for the Diem project\Anon{ and has been maintained by Aptos Labs since 2022; it}{ and} has verified Diem's payment system and critical components of the Aptos blockchain~\cite{ParkZGXGCLC24,AptosFrameworkBook}.

\Section{Hybrid Inference by Example}
\label{sec:example}

Hybrid inference combines two sources of specifications.
The mechanical \WP analysis (\S~\ref{sec:wp}) derives exact contracts but needs loop invariants for code with loops.
The coding agent supplies those invariants and anything else the analysis cannot derive, and it repairs the specification based on verifier feedback.
The prover is connected to the agent through a Model Context Protocol (MCP) service that offers a tool for verification and a tool for \WP inference. 
The agent carries a set of skills describing the Move language, the tools, and the workflow (\S~\ref{sec:skills}).
We walk through one session; the example is written in Move~\cite{MoveOnAptosBook}.

\label{sec:pow}
The example function below computes the power $\mathit{base}^{\mathit{exp}}$ iteratively:

\begin{MoveBoxUser}
  fun pow(base: u64, exp: u64): u64 {
    let result = 1;
    let i = 0;
    while (i < exp) {
      result = result * base;
      i = i + 1;
    }
    result
  }
\end{MoveBoxUser}

Following its skill instructions, the agent first calls \WP. The answer reports the missing loop invariant and, as evidence, unrolls the loop symbolically and shows the values of the loop variables at each of the first loop entries:

\begin{MoveBoxTool}{wp}
WP inferred `vacuous` conditions after this loop without an invariant. ..
  --> sources/pow.move:5:14
  |
5 |       while (i < exp) {
  |              ^
  |
  = loop-invariant evidence 
  = bounded loop-head facts :
      head[0]: head[0].i == 0
               head[0].result == 1
      head[1]: 0 < exp ==> head[1].i == 1
               0 < exp ==> head[1].result == base
      head[2]: 1 < exp ==> head[2].i == 2
               1 < exp ==> head[2].result == base * base
      ..
\end{MoveBoxTool}

From this evidence, the agent derives the loop invariant and introduces a recursive specification helper function for the accumulated result.
It then calls the verification tool to check whether the invariant verifies:

\begin{MoveBoxAgent}
  fun pow(base: u64, exp: u64): u64 {
    ..
    while (i < exp) {
      result = result * base;
      i = i + 1;
    } spec {
      invariant i <= exp;
      invariant result == math_pow(base, i);
    }
    ..   
  }
  // specification helper function
  spec fun math_pow(base: num, exp: num): num {
    if (exp <= 0) 1
    else math_pow(base, exp - 1) * base
  }
\end{MoveBoxAgent}

In the next turn the agent runs the \WP tool again.
With the loop invariant present, \WP derives the complete specification.
However, it marks the abort condition as |sathard|. The marker is set when a derived condition is exact but its satisfiability is expected to be hard for the solver --- here because of the existential quantifier, which states that the computation aborts if the multiplication overflows in some iteration |y|.
The specification language of the Move Prover uses unbounded integers (|num|), so it can express the overflow of the program's |u64| values:

\begin{MoveBoxTool}{wp}
  spec pow { // post and aborts condition for pow
    ensures result == math_pow(base, exp);
    aborts_if [inferred = sathard] exists y: u64:
      math_pow(base, y) * base > MAX_U64
      && (forall x: num: 0 <= x && x <= y
            ==> math_pow(base, x) <= MAX_U64)
      && y < exp
  }
\end{MoveBoxTool}

This walkthrough uses a solver timeout of 5 seconds, shorter than the 40 seconds of the evaluation.
When the agent tries to verify the specification, verification times out with the following diagnosis:

\begin{MoveBoxTool}{verify}
  verification failed:
  error: verification out of resources/timeout
         (timeout set to 5s)
   = timeout analysis replay:
       mixed quantifier and nonlinear arithmetic
   = top quantifier activity:
       quantifier at sources/pow.move:24
         -- 55,793+ observed instantiations
\end{MoveBoxTool}

The diagnosis points to a quantifier: the recursive definition of the helper |math_pow|, which the solver keeps unfolding while trying to prove the quantified |sathard| abort condition.
In the next turn, the agent adds lemmas that bound every partial product by the last one.
With these lemmas, it can restate the abort condition without a quantifier, as an overflow of the last product alone:

\begin{MoveBoxAgent}
  spec pow {
    pragma opaque;
    ensures result == math_pow(base, exp);
    /// The partial products are non-decreasing, so the
    /// loop overflows exactly when its last product does.
    aborts_if
      exp > 0 && math_pow(base, exp - 1) * base > MAX_U64;
  } proof {
    // Bounds every iteration's product by the last one;
    // triggers on the invariant's `math_pow(base, i)`.
    forall k: num {math_pow(base, k)}
      apply math_pow_last_is_max(base, k, exp);
  }
  spec lemma math_pow_last_is_max(
      base: num, k: num, n: num) {
    requires base >= 0 && 0 <= k && k < n;
    ensures math_pow(base, k) * base
              <= math_pow(base, n - 1) * base;
  } proof {
    apply math_pow_mono(base, k + 1, n);
  }
  spec lemma math_pow_mono ...  // left out for brevity
\end{MoveBoxAgent}

\noindent This specification passes verification, and the agent returns it as the result after four turns.

\Section{Weakest Preconditions over Eliminated References}
\label{sec:wp}

Our weakest-precondition analysis~\cite{dijkstra1975wp} derives abort conditions, modifies clauses, and consequence-style postconditions from a function's bytecode.
The analysis itself is the classical one.
To handle Rust-style references, it runs on \MVP's reference-eliminated bytecode, where backward predicate propagation is plain substitution.
To make its output readable, a reverse mapping carries the derived predicates back to the references of the source.

\SubSection{Reference Elimination}
\label{sec:refelim}

Move has Rust-style references: |&T| for shared reads and |&mut T| for exclusive writes, both with a stack-bounded lifetime, and the borrow checker guarantees that a live |&mut| is the only access path to its referent.
\MVP exploits this guarantee to rewrite bytecode into a form in which references are plain values~\cite{DillGPQXZ22}, originally in order to keep the verification conditions first-order and free of a heap model.
Two passes perform the elimination.
Shared references are inlined: |&T| disappears and reads through it become reads of the underlying value.
A mutable reference becomes a pair of the value it currently denotes and a \emph{borrow path} back to its root, which is a local, a |&mut| parameter, or a global resource |R[a]| (\S~\ref{sec:globals}); a write through the reference updates only this value.
Where the lifetime of the reference ends, as determined by a borrow analysis, explicit \emph{write-backs} are inserted along the path up to the first ancestor that is still alive; each rebuilds the parent value with the field or element replaced.
A |&mut| parameter becomes an in/out value parameter: it enters as a plain value and leaves as the value written back last.
Where the root of a reference is not statically unique, as in |if (c) &mut x else &mut y|, the write-back is guarded by an explicit test of which parent the reference was taken from.
Before \WP runs, the bytecode further passes through \MVP's specification instrumentation, which turns each function-level |requires|, |aborts_if|, |ensures|, or |modifies| clause and each loop |invariant| into an |assume| or |assert| at the appropriate program point.
Calls are summarized by the \emph{behavioral predicates} of~\cite{GrieskampEtAl26HO}, |requires_of<f>|, |aborts_of<f>|, |ensures_of<f>|, |modifies_of<f>|, and |result_of<f>|, which reify a callee's contract as predicates over the call site's arguments and result.
At the input to \WP, mutable borrows, writes through references, and write-backs are still instructions of the bytecode, but they operate on values; all other instructions are assignments over plain values, interleaved with |assume| and |assert|.

\SubSection{Global Storage}
\label{sec:globals}

In addition to Rust's reference semantics, Move has \emph{global storage}, a memory indexed by a type and an address.
A resource type |R| and an address |a| identify at most one value, written |R[a]|.
The operations |move_to|, |move_from|, and |exists<R>(a)| publish, remove, and test this value, and |&R[a]| and |&mut R[a]| borrow it.
The prover models global storage as one map from addresses to values per resource type, and the specification language uses the same notation |R[a]|, with |old(R[a])| for the value at function entry.
Since the memory is indexed by type, accesses to different resource types never alias, and accesses to the same type alias only when their addresses coincide.
A function that updates global storage several times passes through several memory states.
The specification language can name such intermediate states with \emph{state labels} and evaluate a condition at a labeled state |S|, written !S |~ e!; |old(R[a])| refers to the state at function entry.

\SubSection{The Predicate Transformer}
\label{sec:transformer}

The analysis is a backward dataflow pass over the control-flow graph.
Its state at a program point is a triple $(E, A, \ell)$: a set $E$ of \emph{ensures} conjuncts that hold on every normal return reached from the point, a set $A$ of \emph{abort} disjuncts under which the function aborts, and a state label $\ell$ (\S~\ref{sec:globals}) naming the memory state the conditions refer to.
The analysis also records which |&mut| parameters and global resources are written between the program point and the function's exits.
At a return, $E = \{\,\mathtt{result}_i = t_i\,\}$ for the returned temporaries and $A = \emptyset$; at an |abort|, $A = \{\mathtt{true}\}$.
Instructions transform the state as follows.
\begin{Itemize}
\item \emph{Assignment and operators.}
  $t := e$ substitutes $e$ for $t$ in $E$ and $A$; an operator that can abort (arithmetic overflow and underflow, division by zero, an index out of range, a cast that truncates, a missing enum variant) also adds its abort predicate to $A$.
\item \emph{References.}
  The central instruction is the write-back, which ends the lifetime of a reference.
  A write-back of $v$ into a parent |x| replaces |x| by $x[f \mapsto v]$, where $f$ is the field, vector index, or map key at which the reference was borrowed from |x|; for a reference returned by a callee, $f$ is a path of fields and the update is nested.
  A write $*r := v$ substitutes $v$ for $r$, and a borrow $r := \,$|&mut x.f| substitutes |x.f| for $r$; together with the write-back, a borrow and the writes through it thus become a functional update of the root.
  A write-back to a global root |R[a]| instead adds |modifies R[a]| and an equation for the new value of |R[a]| at the current state label, and moves $\ell$ to the label before the write-back; the global borrow $r := \,$|&mut R[a]| substitutes |R[a]| for $r$ and adds |!exists<R>(a)| to $A$.
  Since every reference is a path to a named root, no instruction needs a heap.
\item \emph{Calls.}
  A call $t := f(\bar a)$ substitutes |result_of<f>|$(\bar a)$ for $t$ (or the callee's specification function, if $f$ is pure), adds |aborts_of<f>|$(\bar a)$ to $A$, and relates the memory before and after the call through |ensures_of<f>|; when the callee has no abort contract, the abort condition is recorded as \emph{partial} rather than invented.
  Calls are thus summarized modularly: the caller's contract says what $f$ returns without committing to how.
\item \emph{Branches.}
  At a branch $\mathtt{if}\ c\ \mathtt{goto}\ L_1\ \mathtt{else}\ L_2$, the analysis combines the states $(E_1, A_1)$ and $(E_2, A_2)$ computed at the two successors $L_1$ and $L_2$.
  A condition that occurs in both is kept unchanged; any other ensures condition $e \in E_1$ or $e' \in E_2$ becomes $c \Rightarrow e$ or $\neg c \Rightarrow e'$, and any other abort condition $a \in A_1$ or $a' \in A_2$ becomes $c \wedge a$ or $\neg c \wedge a'$.
\item \emph{Specifications.}
  An |assert| $\varphi$ adds $\varphi$ to $E$ and $\neg\varphi$ to $A$; an |assume| $\varphi$ adds $\varphi$ as an antecedent to both.
  This is how the contracts of callees, user-written conditions, and loop invariants enter the derivation.
\item \emph{Loops.}
  Loops are broken at their headers by the cut-point encoding of Barnett and Leino~\cite{BL05}: the back-edge is severed, the loop |invariant| is asserted before and after each iteration, and the loop-modified locations are havocked in between.
  \WP of |havoc x| universally quantifies |x| in $E$ and existentially in $A$, so variables that the invariant does not constrain appear quantified in the contract.
\end{Itemize}
Since the remaining graph is acyclic, the pass visits each block once, in reverse topological order, and the state at entry is the raw contract.
A post-processing step then maps it back to the source (\S~\ref{sec:reverse}) and simplifies it.
Simplification matters more here than in a verifier, where the \WP result is only passed to the solver: the raw predicate of a twenty-line function can have hundreds of terms, and the agent and the user read the simplified one.

\SubSection{Reverse Mapping to Source References}
\label{sec:reverse}

The predicate at entry is stated over the reference-eliminated form, whose in/out parameters have no counterpart in the source signature; the contract the user reads must be stated over the original references.
Three cases arise.
A |&T| parameter needs nothing: the specification language already reads a shared reference as its value.
For a |&mut T| parameter |p|, the specification language reads |p| as the value at exit and |old(p)| as the value at entry. The last value written to |p| becomes |ensures p == e|, where reads of |p| in |e| refer to |old(p)|. In straight-line code, several writes combine into one equation: |*p = *p + 1; *p = *p * 2| yields |ensures p == (old(p) + 1) * 2|. In a loop, |p| is havocked like any loop-modified variable, and the loop invariant has to relate it to |old(p)|.
For a global resource, a borrow of |R[a]| that is written back yields |modifies R[a]| together with an equation on the resource's post-state value, with |old(R[a])| for its value at entry.
Consider the transfer of an amount between two balances:

\begin{MoveBoxUser}
  struct Balance has key { v: u64 }
  fun transfer(from: address, to: address, amount: u64) {
    Balance[from].v -= amount;
    Balance[to].v += amount;
  }
\end{MoveBoxUser}

\noindent Both statements borrow a resource mutably, read and write a field through the reference, and write back.
The derived contract is stated over the resources themselves, and the two borrow roots may coincide, so \WP keeps them apart by a case split on |from == to|:

\begin{MoveBoxTool}{wp}
  spec transfer {
    pragma opaque;
    modifies Balance[from];
    modifies Balance[to];
    aborts_if !exists<Balance>(from);
    aborts_if !exists<Balance>(to);
    aborts_if Balance[from].v < amount;
    aborts_if
      from != to && Balance[to].v + amount > MAX_U64;
    ensures
      from == to ==> Balance[from] == old(Balance[from]);
    ensures from != to
      ==> Balance[from].v == old(Balance[from]).v - amount;
    ensures from != to
      ==> Balance[to].v == old(Balance[to]).v + amount;
  }
\end{MoveBoxTool}

\noindent When a function updates global memory in several steps, an equation on the final memory is not enough: the abort condition of the second |move_to| below depends on the memory after the first.
The state labels of the transformer therefore appear in the contract, naming the intermediate memory states at which a global update happens; a condition is evaluated at a labeled state, !S |~ e!, or over a range of states, !S1..S2 |~ e!:

\begin{MoveBoxUser}
  fun split_balance(
      dst1: &signer, dst2: &signer, src: address): u64 {
    let Balance { v } = move_from<Balance>(src);
    let half = v / 2;
    move_to(dst1, Balance { v: half });
    move_to(dst2, Balance { v: v - half });
    half
  }
\end{MoveBoxUser}
\begin{MoveBoxTool}{wp}
  spec split_balance {
    pragma opaque;
    modifies Balance[src];
    modifies Balance[signer::address_of(dst1)];
    modifies Balance[signer::address_of(dst2)];
    ensures result == old(Balance[src]).v / 2;
    ensures ..S1 |~ remove<Balance>(src);
    ensures S1..S2 |~
      publish<Balance>(signer::address_of(dst1),
        Balance { v: old(Balance[src]).v / 2 });
    ensures S2.. |~
      publish<Balance>(signer::address_of(dst2),
        Balance { v: old(Balance[src]).v
                     - old(Balance[src]).v / 2 });
    aborts_if !exists<Balance>(src);
    aborts_if
      S1 |~ exists<Balance>(signer::address_of(dst1));
    aborts_if
      S2 |~ exists<Balance>(signer::address_of(dst2));
  }
\end{MoveBoxTool}

\noindent |S1| names the memory between the |move_from| and the first |move_to|, |S2| the memory between the two |move_to|s; |publish<R>(a, x)| states that resource |R| at |a| equals |x| over the range, |remove<R>(a)| that none is present, and |exists<R>(a)| is the usual existence predicate.
Without the labels, the memory versions the eliminated form threads through the function could be read back only as a single, weaker equation on the final memory.

\SubSection{Diagnostics for the Agent}
\label{sec:diagnostics}

The contract that \WP derives is only as complete as its inputs.
A loop without an invariant leaves a gap, and so does a callee without an abort contract.
Rather than silently emitting an incomplete contract, the tool reports both.
After a havocked loop, a condition that quantifies over everything the loop touched carries no information; such \emph{vacuous} conditions are dropped and the loop is reported with \emph{loop-invariant evidence} --- the facts that hold at the first few loop heads on each path, computed by bounded unrolling of the same transformer, as in the |head[k]| listing of \S~\ref{sec:example}.
A condition that is exact but existentially quantified, typically an abort condition that says \emph{some} iteration overflows, is kept and marked |sathard|, since its satisfiability will be expensive for the solver.
An abort condition that could not account for a callee is marked \emph{partial}.
Every emitted condition carries an |[inferred]| marker, so that re-runs replace their own output and never disturb user-written specifications (the listings in this paper omit the marker, except where it carries |sathard|).
The agent's skills are written against these markers: they identify the loop that needs an invariant, the condition to simplify, and the callee to specify first.

\SubSection{Inference without a Heap Model}
\label{sec:advantages}

Reference elimination was introduced to make \emph{verification} of Move tractable~\cite{DillGPQXZ22}: without a heap, the verification conditions stay first-order and the solver needs no frame reasoning.
A predicate transformer over programs with references has to carry a model of locations: |*r| means the contents of whatever |r| points to, an assignment through one reference may change the value read through another, and the transformer needs either an explicit heap with separation or ownership reasoning to keep the substitutions sound~\cite{RustHorn,Aeneas}.
Its output is then stated over that model --- heap fragments, prophecy variables, ownership predicates --- and has to be translated before a user can read it.
On the eliminated form none of this arises.
Every reference is a path to a named root, so substitution is exact by construction; the only aliasing that survives is between global resources whose addresses may coincide, and it surfaces as an explicit condition such as |from == to| above.
The derived predicates mention only parameters, results, fields, and resources, which is the vocabulary of the specification language, so the reverse mapping of \S~\ref{sec:reverse} reduces largely to renaming.
And because inference and verification run on the same pipeline, every inferred contract is checked under the semantics that produced it; the instrumented contracts of callees and the loop invariants are available in-line to both.
The whole analysis is about nine thousand lines of Rust on top of the existing pipeline, by far the largest part of it being the simplification of predicates, and it adds no new semantics to the Move Prover.


\Section{Agentic Inference}
\label{sec:skills}

The agentic side of the workflow is a plugin for a coding agent: skills that instruct it, hooks that check generated Move as it is written, and the Move toolchain exposed via a local MCP server. The hooks run after every edit of a Move file: they parse the file, check its specification expressions, reject deprecated Move~1 constructs, and format the source, so that syntactic slips are caught before a verification run is spent on them. The plugin is rendered per tactic from one source; the setup is at~\Anon{\cite{MoveFlow}}{\cite{MoveFlowAnon}}, and the evaluation of \S~\ref{sec:evaluation} runs the rendered plugin under the Codex CLI, which receives the skills inline with the task and the tools through the MCP server. We summarize the design.

\Paragraph{Inference tactics}
Inference runs in one of three \emph{tactics}, which differ in whether the \WP tool is available and who decides when to use it.
\begin{Itemize}
\item \emph{agent-only:} \WP is unavailable. The contract and any loop invariants are derived directly from the implementation and the contracts of its dependencies, checked as the work proceeds.
\item \emph{hybrid-guided:} the workflow of \S~\ref{sec:example}, prescribed as a fixed order --- run \WP over the scope, take one warned function and write an invariant from the loop-invariant evidence (\S~\ref{sec:diagnostics}), rerun \WP on that function alone until the warning is gone, then check.
\item \emph{hybrid-flexible:} \WP is available as a tool, and the agent decides whether and when to call it and how to combine it with its own reasoning.
\end{Itemize}

\Paragraph{Skills}
A tactic selects the \emph{plan} --- the sequence of tasks and which of them consult \WP. All other guidance is shared reference material, identical in all three tactics: the specification language, the rules for writing and editing specs, the verification sub-workflow, and the toolchain's capabilities and limits. Only the description of the \WP tool itself is withheld where the tool is absent; the underlying idea of backward reasoning, and what a loop without an invariant costs, are background that every tactic receives.

The most relevant areas of guidance are:
\begin{Itemize}
\item \emph{Loop invariants:} when to introduce recursive helper functions for iterative quantities, and when to prefer state-level invariants on resources --- assumed by the prover at every call site, including each loop iteration --- over per-loop inductive arguments.
\item \emph{Timeout resolution:} a hierarchy of escalations --- strengthen state-level invariants, then introduce spec helpers, then lemmas, then explicit proof scripts --- with a strict prohibition on weakening abort or postconditions to silence a timeout, since this gives up a real property only to make the proof succeed.
\end{Itemize}


\Section{Evaluation}
\label{sec:evaluation}

The evaluation addresses three research questions: does adding \WP to an agent change the completeness of the inferred contracts (RQ1), does it lower the cost of inference, and on which targets (RQ2), and with \WP, does a cheaper mid-tier model match the results of a frontier model (RQ3)?
\S~\ref{sec:eval-arms} answers RQ1 and RQ2 by holding the model fixed and comparing the three tactics, \textit{agent-only}, \textit{hybrid-guided}, and \textit{hybrid-flexible}, under the same conditions.
\S~\ref{sec:eval-models} answers RQ3 by repeating the comparison under a frontier model and a mid-tier model of the same family.
Both comparisons share the corpus, the protocol, the scoring, and the cost accounting described next.

\SubSection{Setup}
\label{sec:eval-setup}

\Paragraph{Corpus}
The corpus has 26 targets from 12 modules.
Twenty-four are extracted from public Move code of the Aptos framework at a pinned commit: ten from its experimental trading modules, thirteen from modules for accounts, authentication, and rate and stake limits, and one from the Move standard library.
The other two are written for the corpus.
Function bodies are copied unchanged except for a few reductions documented in the module headers, and the package includes everything the targets depend on.
No target has a specification upstream.
Thirteen targets have a loop that needs an invariant.
Each target has a hand-written reference specification that verifies within 20 seconds, and six \emph{mutants}: patches to the implementation, each violating one obligation the contract must capture, and each shown to be killed by the reference.
Each inference run starts from the package with the reference contracts of every function the target calls; the contracts of the target and of all other functions are withheld, since a caller's contract can restate what the target does.

\Paragraph{Tactics, models, and cells}
The three tactics of \S~\ref{sec:skills} differ only in the rendered plugin: the reference material is byte-identical, and in agent-only the \WP tool is absent rather than discouraged.
We run them under two models of one family through the same runtime, the Codex CLI at high effort: Sol~5.6, the frontier model, and Terra~5.6, the mid-tier model, priced at about half of Sol's list price.
A \emph{cell} is one run of one tactic on one target; a \emph{round} is the set of cells of one model.
With 26 targets, 3 tactics, and 4 replicates, each round has 312 cells.
Each session has a budget of 100k output tokens, one hour of wall time, and 40 seconds of solver time per verification condition.

\Paragraph{Scoring}
A cell reaches \emph{operational success} when a fresh checker process outside the agent's workspace confirms that the package compiles, the requested scope verifies, no weakening construct is present (|pragma verify = false|, partial abort contracts, vacuous postconditions, invented preconditions), the bytecode is unchanged, edits stay within the allowed scope, and the required contract categories are present.
The agent can run the same check on a candidate contract during the session (a \emph{candidate check}); a check is rejected when one of these criteria fails.
The agent's own claim is never a score, and the agent receives only this acceptance feedback: the mutants are withheld until generation has ended, and a contract is never handed back for strengthening.
Operational success is necessary but weak: a vaguer contract passes it more easily.
Completeness, postconditions as strong as the behavior, has no deductive check in general; \WP guarantees it only given complete loop invariants and callee contracts.
The second level is therefore \emph{mutation adequacy}: after the round, each withheld mutant is applied to the implementation, and the prover is rerun against the inferred specification.
A cell is \emph{complete} when its specification kills all six mutants and \emph{failed} when one survives; a cell in which a mutant reached no verdict within the solver's time limit is \emph{unmeasured}.
All 624 cells reached operational success, so the outcome of a cell is decided by the withheld mutants alone.

\Paragraph{Cost accounting}
Costs are in US dollars at each model's published list prices, computed from the recorded token counters of every model request; startup requests that do no task work are excluded.
Per million tokens, Sol~5.6 is \$4 uncached input, \$0.40 cached input, and \$20 output; Terra~5.6 is \$2, \$0.20, and \$12.
Thinking counts as output for both.
Means are taken first over the four replicates of a target and then over targets, and the 95\% intervals are percentile bootstraps over the 26 targets (10,000 resamples).
Differences in completeness are tested for two preregistered contrasts, hybrid-flexible against agent-only and hybrid-guided against hybrid-flexible, with randomization tests blocked by target and replicate and Holm-adjusted across the two.

\SubSection{Comparison of Tactics (RQ1, RQ2)}
\label{sec:eval-arms}

Table~\ref{tab:summary} summarizes both rounds, Fig.~\ref{fig:cells} shows every cell, and Tables~\ref{tab:sol} and~\ref{tab:terra} in the appendix give every target.

\begin{table}[t]
\centering
\caption{Both rounds, per tactic: mean cost in US dollars per cell, the hybrid/agent-only cost ratio with its 95\% bootstrap interval, complete, failed, and unmeasured cells, the share of cells whose first candidate check was accepted, mean input/output tokens in thousands, and mean wall time per cell in seconds.}
\label{tab:summary}
\scriptsize
\setlength{\tabcolsep}{2.2pt}
\begin{tabular}{llrlrrrrrr}
\toprule
model & tactic & cost & vs. agent [95\% CI] & compl. & failed & unm. & 1st & I/O k & wall s \\
\midrule
Sol~5.6 & agent-only & \$0.353 & --- & 103/104 & 1 & 0 & 62\% & 333/3.8 & 121 \\
 & hybrid-guided & \$0.304 & 0.86 [0.66, 1.09] & 98/104 & 5 & 1 & 68\% & 288/3.5 & 126 \\
 & hybrid-flexible & \$0.306 & 0.87 [0.74, 0.99] & 98/104 & 2 & 4 & 71\% & 280/3.3 & 115 \\
\midrule
Terra~5.6 & agent-only & \$0.202 & --- & 97/104 & 5 & 2 & 37\% & 367/3.9 & 153 \\
 & hybrid-guided & \$0.161 & 0.79 [0.59, 1.02] & 97/104 & 6 & 1 & 63\% & 289/3.1 & 148 \\
 & hybrid-flexible & \$0.181 & 0.89 [0.73, 1.05] & 98/104 & 6 & 0 & 65\% & 335/3.4 & 143 \\
\bottomrule
\end{tabular}

\end{table}

\begin{figure}[t]
\centering
\newcommand{\cellplot}[1]{%
  \addplot[forget plot, only marks, scatter, scatter src=explicit symbolic, mark size=1.25pt,
    scatter/classes={
      ok0={mark=*, ModelSol}, fail0={mark=o, ModelSol, thick}, unm0={mark=x, CellMuted, thick, mark size=1.8pt},
      ok1={mark=*, ModelTerra}, fail1={mark=o, ModelTerra, thick}, unm1={mark=x, CellMuted, thick, mark size=1.8pt}}]
    table[x index=1, y expr={\thisrowno{0} + (\thisrowno{2} - 0.5) * 0.36}, meta index=3] {#1};
  \draw[CellAxis] (axis cs:0.04,{\CellLoopFree - 0.5}) -- (axis cs:2.2,{\CellLoopFree - 0.5});}
\begin{tikzpicture}
\begin{groupplot}[
  group style={group size=3 by 1, horizontal sep=2.5mm, yticklabels at=edge left},
  scale only axis, width=2.5cm, height=7.4cm,
  xmode=log, xmin=0.04, xmax=2.2, log ticks with fixed point,
  xtick={0.1,0.3,1}, xticklabels={0.1,0.3,1},
  minor xtick={0.05,0.2,0.5,2}, minor tick style={draw=none},
  ymin=-0.6, ymax=25.6, ytick={0,...,25}, y dir=reverse,
  ymajorgrids, grid style={CellGrid, thin},
  axis line style={CellAxis}, tick style={CellAxis},
  tick label style={font=\scriptsize},
  title style={font=\small, yshift=-1ex}]
\nextgroupplot[title=agent-only, yticklabels/.expanded={\CellTaskLabels}, yticklabel style={font=\ttfamily\tiny}]
\cellplot{tables/agent_only-cells.dat}
\nextgroupplot[title=hybrid-guided, xlabel={cost per cell (US\$)}, xlabel style={font=\scriptsize},
  legend style={at={(0.5,-0.09)}, anchor=north, font=\scriptsize, legend columns=-1, draw=none, /tikz/every even column/.append style={column sep=0.6em}},
  legend image post style={mark size=1.4pt}]
\cellplot{tables/hybrid_guided-cells.dat}
\addlegendimage{only marks, mark=*, ModelSol} \addlegendentry{Sol~5.6}
\addlegendimage{only marks, mark=*, ModelTerra} \addlegendentry{Terra~5.6}
\addlegendimage{only marks, mark=o, black, thick} \addlegendentry{mutant survived}
\addlegendimage{only marks, mark=x, CellMuted, thick, mark size=2pt} \addlegendentry{unmeasured}
\nextgroupplot[title=hybrid-flexible]
\cellplot{tables/hybrid_flexible-cells.dat}
\end{groupplot}
\end{tikzpicture}
\caption{Every cell of both rounds: cost in US dollars on a log scale, one panel per tactic, one row per target, and within each row the four Sol~5.6 replicates above the four Terra~5.6 replicates. The 13 targets above the line have no loop that needs an invariant. A filled dot is a complete cell, a hollow one a cell a withheld mutant survived, a cross an unmeasured cell.}
\label{fig:cells}
\end{figure}

\Paragraph{RQ1: Completeness}
Measured completeness is about the same under all three tactics: agent-only completes 103 of 104 cells under Sol~5.6 and 97 under Terra~5.6, the hybrids 98 and 98, and 97 and 98, and under both models both preregistered contrasts have Holm $p = 1.00$.
This is what one would expect: the loop invariants and specification helpers that decide the hard targets have to be found by the agent under every tactic.
All failures but one Terra~5.6 agent-only cell occur on three loop targets.
|partition|, an in-place partition, is the one target whose contract needs a permutation argument proved with lemmas; it fails under every tactic, and in every failure a mutant that overwrites an element instead of swapping it survives: the contract does not state that the result is a rearrangement of the input.
On the two scans |upsert_provider_jwks| and |cancel_at_price_level|, which leave the loop early, only hybrid cells fail (5 under Sol~5.6, 6 under Terra~5.6): the agent added invariants for the scanned prefix, accepted the quantified contract that \WP derived from them once it verified, and a withheld mutant on the exit path survived.
The hybrids do improve the first attempt: the share of cells whose first candidate check is accepted rises from 62\% to 68--71\% under Sol~5.6 and from 37\% to 63--65\% under Terra~5.6, and rejected checks drop from 66 to 40--60 and from 117 to 66--82.

\Paragraph{RQ2: Cost}
Under both models the hybrids cost less than agent-only: 0.86$\times$ and 0.87$\times$ under Sol~5.6 and 0.79$\times$ and 0.89$\times$ under Terra~5.6 (hybrid-guided, hybrid-flexible).
Over all targets, only the interval of Sol's hybrid-flexible excludes parity, but the targets split into two clear groups.
On the 13 targets without a loop that needs an invariant, \WP derives the complete contract and the agent only has to check it: the hybrids cost 0.36--0.60$\times$ agent-only, and every interval lies below 0.75.
On the 13 loop targets they cost 1.09--1.22$\times$: the agent still has to find the invariant, and the quantified \WP output adds to the context it re-reads on every turn --- on these targets the hybrids read 1.2--1.4$\times$ the input of agent-only (415--513k against 357--385k tokens per cell), on the loop-free ones 0.24--0.47$\times$.
Overall the hybrids read less input than agent-only (288k and 280k against 333k tokens per cell under Sol~5.6, 289k and 335k against 367k under Terra~5.6) and write less output.

\SubSection{Comparison of Models (RQ3)}
\label{sec:eval-models}

Terra~5.6 costs 0.57$\times$ Sol~5.6 under agent-only and 0.53$\times$ and 0.59$\times$ under the hybrids, with all intervals within [0.45, 0.65], and it is cheaper on 25 or 26 of the 26 targets under every tactic.
These ratios are close to the price ratio of 0.5 for input and 0.6 for output: the two models spend about the same number of tokens per cell, Terra~5.6 up to a fifth more input, and Terra~5.6 is slower (143--153 against 115--126 seconds).
Completeness differs on one target.
Outside |partition|, Terra~5.6 fails one agent-only cell and, under the hybrids, the same two scans as Sol~5.6; on the partition it fails 10 of 12 cells, Sol~5.6 3 of 12.
The frontier model's advantage therefore lies in the lemma-based permutation argument; on the loop invariants, compositions, and global state that make up the rest of the corpus the two models are even.
The one target on which Terra~5.6 costs about as much as Sol~5.6 is |select|, whose contract needs a recursive specification function over a function value; there it writes 1.5--2.0$\times$ Sol's output.
With \WP, Terra's first-check acceptance (63--65\%) approaches Sol's (68--71\%), and Terra~5.6 under hybrid-guided completes 97 of 104 cells at 46\% of the cost of Sol~5.6 under agent-only, which completes 103.


\Section{Discussion}
\label{sec:discussion}

\Paragraph{Where \WP helps}
On targets without loops, \WP needs no help: it derives the complete contract, and the
agent only checks it, at 36--60\% of the cost of writing the contract directly. Half of
this corpus is of this kind, including four of the five compositions over callee
contracts and eight of the nine targets with global state. On loop targets the agent still has to find the
invariant, and \WP's quantified output adds context that is re-read on every turn, so
the hybrids cost slightly more there. With \WP, the first candidate check is accepted more
often, most for the mid-tier model (from 37\% to 63--65\% of cells).

\Paragraph{Where it does not}
The partition needs a permutation argument that \WP cannot supply, and every tactic fails
on it, the mid-tier model more often. The two scans that leave their loop early fail only
in the hybrid tactics: the agent accepted the contract that \WP derived from its prefix
invariants once it verified, and a mutant on the exit path survived. A derived contract
that verifies can thus still be incomplete, and the tool's diagnostics, which report a
missing invariant, do not detect this case.

\Paragraph{Mid-tier and frontier models}
Both models spend about the same number of tokens per cell, so the cost of the mid-tier
model follows its price. Apart from the permutation target, it handles the loop
invariants, compositions, and global state of the corpus as well as the frontier model.
We expect the cost advantage of \WP to extend to cheaper models: for a local model, the
context \WP adds costs latency rather than money, and a model trained to read \WP output
should need fewer rounds with the tool.

\Paragraph{Carrying over to prophecies}
The reverse mapping of \S~\ref{sec:reverse} relies on the elimination being
\emph{path-based}: an eliminated reference is still a path to a location the source can
name, so the derived predicate has a source-level spelling. A \emph{prophecy}-based
elimination in the style of RustHorn and Creusot~\cite{RustHorn,Creusot22}, in which a mutable
reference is a pair of its current value and a prophecy variable for its final value,
pinned when the borrow ends, has been designed for \MVP but not implemented; it would
need no borrow paths and would handle borrows through function values, which the
path-based model cannot. A contract is model-agnostic, so inference could keep running
on the path-based form regardless. Whether \WP could run on the prophecy form itself
and still be carried back is open. For a |&mut| parameter the prophecy \emph{is} the
post-state value and has the spelling |p|; for a reborrow through a function value the
final value of the borrow has no source-level counterpart at all, and a predicate over it
would have to be existentially closed or restated through the callee's contract. We do
not know yet whether that always yields a readable contract.

\Paragraph{Threats to validity}
The reference specifications, the mutants, and the skills were written by the authors,
who also built \WP; the agent-only tactic receives the same reference material, and the
mutants are the same for every tactic, but a different corpus or different skills could
favor a tactic differently. Six mutants per target probe six obligations, so a contract
that kills them all may still be weaker than the reference in ways the mutants do not
test. Four replicates per target and tactic bound what the comparison can resolve: the agent is not
deterministic, and differences of a few cells per tactic lie within that noise. Costs are
list prices of one provider under one runtime, and the targets are functions of one
codebase under two models of one family.

\Paragraph{Open questions}
Whether \WP changes completeness is open. What we measured is about equal across
tactics, and that is not surprising, since the loop invariants and specification helpers
that decide the hard targets have to be found by the agent in every tactic; but the
measure is coarse, and a difference small enough to hide behind six mutants and four
replicates would still matter on a larger corpus. The scale of inference is open as well:
the targets are functions, not modules. We plan to apply inference to the parts of the
Aptos framework~\cite{AptosFrameworkBook} that already carry hand-written specifications
and to compare the inferred contracts with them --- a comparison this corpus cannot make,
since its targets were selected for having no specification.

\Section{Related Work}
\label{sec:related}

Specification inference predates large language models --- Daikon~\cite{DaikonTSE01} learned likely invariants from program traces, Houdini~\cite{Houdini01} kept the candidate annotations a verifier could not refute, and bi-abduction~\cite{BiAbduction09} inferred shape preconditions and postconditions compositionally --- but these techniques stayed tied to particular property classes and source languages, and the advent of LLMs has revived interest in the area.
Pei et al.~\cite{PeiEtAl23} showed that fine-tuned LLMs can predict program invariants statically at quality competitive with dynamic analysis.
Kamath et al.~\cite{KamathEtAl23} pair an LLM that \emph{proposes} loop invariants with a symbolic checker that \emph{validates} them, retrying on failure --- a propose-and-validate loop that recurs in much of the subsequent work.
For Verus, AlphaVerus~\cite{AlphaVerus} bootstraps verified code by self-improving translation with tree search over verifier feedback, and VeriStruct~\cite{VeriStruct26} synthesizes Verus-style specifications and proofs for Rust functions, modules, and data structures through planner-orchestrated phases.
AutoSpec~\cite{AutoSpec24} generates ACSL specifications for C with an LLM guided by static analysis of the call graph and validated by Frama-C, and SpecGen~\cite{SpecGen25} generates JML specifications for Java in a conversation with the verifier followed by mutation of the candidates.
For smart contracts in Solidity, PropertyGPT~\cite{PropertyGPT25} retrieves human-written Certora properties as input examples for specification generation.
Most directly related to our work, MSG~\cite{MSG25} targets the same Move specification language through a skill-based agentic design that consumes verifier feedback beyond a binary success/failure verdict.
Across these systems the LLM remains responsible for the entire specification; the verifier serves as a binary oracle or, in MSG, as a feedback channel, but never as a contributor of mechanically derived content.
A separate line of work, exemplified by DafnyPro~\cite{DafnyPro26} and AutoVerus~\cite{AutoVerus25}, tackles the complementary problem of synthesizing the auxiliary proof annotations a verifier needs when the specification is already given. Our plugin can also synthesize proof hints (\S~\ref{sec:pow}), but this is not the focus of this paper.

Reference elimination, the transformation underlying our \WP analysis, was introduced in~\cite{DillGPQXZ22}, and equivalents exist for Rust.
Aeneas~\cite{Aeneas} compiles Rust to a pure functional model in a target prover (Lean, F*, Coq, HOL4) via forward and backward functions, while RustHorn~\cite{RustHorn} encodes a mutable borrow as a pair of its current value and a prophecy for its value at the end of the borrow, reducing verification to constrained Horn clauses.
Neither infers specifications --- in Aeneas the user writes them in the target prover, against the translated model --- and both consume the reference-free form only in the forward direction.
Our use of the transformation is the opposite: \WP runs on the reference-free form, and the conditions it derives are carried \emph{back} through the elimination.
To our knowledge this inversion has not been described before.
Boogie~\cite{boogie,boogieIVL}, the intermediate verification language that \MVP translates to, also computes weakest preconditions of the procedures it is given~\cite{BL05} and passes them to the SMT solver; the difference is that Boogie's \WP is not intended for external consumption, whereas ours has to be read, which is reflected in the reverse mapping and in the effort spent on simplification.

Two aspects thus distinguish our setup from the work above.
First, to our knowledge, no prior LLM-based specification inference system combines the LLM with a sound analysis that itself contributes specification content.
\WP itself produces a substantial part of the inferred specification --- abort conditions, modifies clauses, and consequence-style postconditions --- so the AI is left only with what \WP cannot recover, and the two streams are validated together by the Move Prover.
Second, the workflow is delivered as a plugin to a general-purpose, interactive coding agent via skills, edit-time hooks, and an MCP server, rather than as a custom-built tool: the agent participates in the loop and the user can intervene at any point.

\Section{Conclusion}
\label{sec:conclusion}

We have presented a specification inference approach for the Move Prover that runs \WP on reference-eliminated bytecode, where backward propagation is substitution and the derived contracts can be mapped back to the references of the source, and we have combined it with a coding agent that supplies what \WP cannot: loop invariants, restated abort conditions, and proof hints. The prover checks the combined result.
Reference elimination was introduced for verification; this work shows that it serves inference as well, and the approach should carry over to verifiers for Rust-like languages that eliminate references along paths.
On 26 targets, the agent alone already completes most cells, and \WP lowers the cost wherever it derives the contract by itself.
We expect local and specialized models, for which the additional context costs least, to benefit further from \WP.


\subsubsection*{Data Availability Statement.}
The corpus is published with the inference plugin at~\Anon{\cite{MoveFlow}}{\cite{MoveFlowAnon}}: the package extracted from a pinned commit of the Aptos repository, the reference specifications, and the mutants.
The round archives behind every table and figure --- per-cell costs, token counters, wall times, and mutant outcomes --- and the scripts that produce them from the archives are published there as well; \Anon{\cite{SpecInferenceCorpus}}{\cite{MoveFlowAnon}} explains how to verify the corpus, re-score a round, and rerun the experiment.

\subsubsection*{Use of AI Tools.}
Parts of the text of this paper were drafted and edited with a coding agent\Anon{ (Claude Code)}{}; the authors reviewed and take responsibility for all content.

\bibliography{biblio}

\begin{thebibliography}{10}
\providecommand{\url}[1]{\texttt{#1}}
\providecommand{\urlprefix}{URL }
\providecommand{\doi}[1]{https://doi.org/#1}

\bibitem{AlphaVerus}
Aggarwal, P., Parno, B., Welleck, S.: {AlphaVerus}: Bootstrapping formally
  verified code generation through self-improving translation and treefinement.
  In: Proceedings of the 42nd International Conference on Machine Learning
  ({ICML} 2025). Proceedings of Machine Learning Research, vol.~267 (2025)

\bibitem{AptosFrameworkBook}
{Aptos Labs}: {The Aptos Framework Book}.
  \url{https://aptos-labs.github.io/framework-book/} (2023), accessed:
  2026-05-09

\bibitem{MoveOnAptosBook}
{Aptos Labs}: {The Move on Aptos Book}.
  \url{https://aptos-labs.github.io/move-book/} (2023), accessed: 2026-05-09

\bibitem{MoveFlow}
{Aptos Labs}: {MoveFlow}: {AI}-assisted {Move} smart contract development.
  \url{https://github.com/aptos-labs/aptos-core/tree/c1737fde43/aptos-move/flow}
  (2026), commit \texttt{c1737fde43}, accessed 2026-10-06

\bibitem{SpecInferenceCorpus}
{Aptos Labs}: Reproducing the corpus-v4 specification-inference results.
  \url{https://github.com/aptos-labs/aptos-core/blob/c1737fde43/aptos-move/flow/evaluation/spec-inference/corpus-v4/REPRODUCE.md}
  (2026), commit \texttt{c1737fde43}, accessed 2026-10-06

\bibitem{DafnyPro26}
Banerjee, D., Bouissou, O., Zetzsche, S.: {DafnyPro}: {LLM}-assisted automated
  verification for {Dafny} programs (2026),
  \url{https://arxiv.org/abs/2601.05385}, dafny Workshop, {POPL} 2026

\bibitem{boogie}
Barnett, M., Chang, B.Y.E., DeLine, R., Jacobs, B., Leino, K.R.M.: Boogie: A
  modular reusable verifier for object-oriented programs. In: International
  Symposium on Formal Methods for Components and Objects. pp. 364--387.
  Springer (2005)

\bibitem{BL05}
Barnett, M., Leino, K.R.M.: Weakest-precondition of unstructured programs. In:
  Proceedings of the 6th ACM SIGPLAN-SIGSOFT Workshop on Program Analysis for
  Software Tools and Engineering. pp. 82---87. Association for Computing
  Machinery, New York, NY, USA (2005). \doi{10.1145/1108792.1108813}

\bibitem{BiAbduction09}
Calcagno, C., Distefano, D., O'Hearn, P.W., Yang, H.: Compositional shape
  analysis by means of bi-abduction. In: Proceedings of the 36th {ACM}
  {SIGPLAN-SIGACT} Symposium on Principles of Programming Languages ({POPL}
  2009). pp. 289--300. {ACM} (2009)

\bibitem{CousotHalbwachs78}
Cousot, P., Halbwachs, N.: Automatic discovery of linear restraints among
  variables of a program. In: Proceedings of the 5th {ACM} {SIGACT-SIGPLAN}
  Symposium on Principles of Programming Languages ({POPL} '78). pp. 84--96.
  {ACM} (1978). \doi{10.1145/512760.512770}

\bibitem{Creusot22}
Denis, X., Jourdan, J.H., March{\'e}, C.: Creusot: A foundry for the deductive
  verification of {Rust} programs. In: Formal Methods and Software Engineering
  (ICFEM 2022). LNCS, vol. 13478, pp. 90--105. Springer (2022).
  \doi{10.1007/978-3-031-17244-1_6}

\bibitem{dijkstra1975wp}
Dijkstra, E.W.: Guarded commands, nondeterminacy and formal derivation of
  programs. Communications of the ACM  \textbf{18}(8),  453--457 (1975).
  \doi{10.1145/360933.360975}

\bibitem{DillGPQXZ22}
Dill, D.L., Grieskamp, W., Park, J., Qadeer, S., Xu, M., Zhong, J.E.: Fast and
  reliable formal verification of smart contracts with the move prover. In:
  Tools and Algorithms for the Construction and Analysis of Systems ({TACAS}
  2022). Lecture Notes in Computer Science, vol. 13243, pp. 183--200. Springer
  (2022). \doi{10.1007/978-3-030-99524-9_10}

\bibitem{DaikonTSE01}
Ernst, M.D., Cockrell, J., Griswold, W.G., Notkin, D.: Dynamically discovering
  likely program invariants to support program evolution. {IEEE} Transactions
  on Software Engineering  \textbf{27}(2),  99--123 (2001).
  \doi{10.1109/32.908957}

\bibitem{Houdini01}
Flanagan, C., Leino, K.R.M.: Houdini, an annotation assistant for {ESC/Java}.
  In: Formal Methods Europe ({FME} 2001). Lecture Notes in Computer Science,
  vol.~2021, pp. 500--517. Springer (2001)

\bibitem{MSG25}
Fu, Y.F., Xu, M., Kim, T.: Agentic specification generator for {Move} programs.
  In: Proceedings of the 40th {IEEE/ACM} International Conference on Automated
  Software Engineering ({ASE} 2025). pp. 1286--1298 (2025).
  \doi{10.1109/ASE63991.2025.00110}

\bibitem{GrieskampEtAl26HO}
Grieskamp, W., Zhang, T., Kashyap, V., Silverman, J.: Formal verification of
  imperative first-class functions in {Move}. CoRR  \textbf{abs/2605.10007}
  (2026), \url{https://arxiv.org/abs/2605.10007}, preprint; to appear in
  {FMCAD} 2026

\bibitem{Aeneas}
Ho, S., Protzenko, J.: {Aeneas}: Rust verification by functional translation.
  Proc. ACM Program. Lang.  \textbf{6}({ICFP}),  116:1--116:31 (2022).
  \doi{10.1145/3547647}

\bibitem{KamathEtAl23}
Kamath, A., Senthilnathan, A., Chakraborty, S., Deligiannis, P., Lahiri, S.K.,
  Lal, A., Rastogi, A., Roy, S., Sharma, R.: Finding inductive loop invariants
  using large language models (2023), \url{https://arxiv.org/abs/2311.07948}

\bibitem{boogieIVL}
Leino, K.R.M., R{\"u}mmer, P.: A polymorphic intermediate verification
  language: Design and logical encoding. In: Esparza, J., Majumdar, R. (eds.)
  TACAS. pp. 312--327. Springer Berlin Heidelberg, Berlin, Heidelberg (2010)

\bibitem{PropertyGPT25}
Liu, Y., Xue, Y., Wu, D., Sun, Y., Li, Y., Shi, M., Liu, Y.: {PropertyGPT}:
  {LLM}-driven formal verification of smart contracts through
  retrieval-augmented property generation. In: Network and Distributed System
  Security Symposium ({NDSS} 2025) (2025). \doi{10.14722/ndss.2025.241357}

\bibitem{SpecGen25}
Ma, L., Liu, S., Li, Y., Xie, X., Bu, L.: {SpecGen}: Automated generation of
  formal program specifications via large language models. In: Proceedings of
  the 47th {IEEE/ACM} International Conference on Software Engineering ({ICSE}
  2025). {IEEE} (2025)

\bibitem{RustHorn}
Matsushita, Y., Tsukada, T., Kobayashi, N.: {RustHorn}: {CHC}-based
  verification for {Rust} programs. ACM Trans. Program. Lang. Syst.
  \textbf{43}(4),  1--54 (2021). \doi{10.1145/3462205}

\bibitem{ParkZGXGCLC24}
Park, J., Zhang, T., Grieskamp, W., Xu, M., Giacomo, G.D., Chen, K., Lu, Y.,
  Chen, R.: {Securing Aptos Framework with Formal Verification}. In: 5th
  International Workshop on Formal Methods for Blockchains ({FMBC} 2024). Open
  Access Series in Informatics (OASIcs), vol.~118, pp. 9:1--9:16. Schloss
  Dagstuhl -- Leibniz-Zentrum f{\"u}r Informatik (2024).
  \doi{10.4230/OASIcs.FMBC.2024.9}

\bibitem{PeiEtAl23}
Pei, K., Bieber, D., Shi, K., Sutton, C., Yin, P.: Can large language models
  reason about program invariants? In: Proceedings of the 40th International
  Conference on Machine Learning ({ICML} 2023). Proceedings of Machine Learning
  Research, vol.~202, pp. 27496--27520. {PMLR} (2023),
  \url{https://proceedings.mlr.press/v202/pei23a.html}

\bibitem{VeriStruct26}
Sun, C., Sun, Y., Amrollahi, D., Zhang, E., Lahiri, S.K., Lu, S., Dill, D.L.,
  Barrett, C.: {VeriStruct}: {AI}-assisted automated verification of
  data-structure modules in {Verus}. In: Tools and Algorithms for the
  Construction and Analysis of Systems ({TACAS} 2026). Lecture Notes in
  Computer Science, Springer (2026), \url{https://arxiv.org/abs/2510.25015},
  arXiv:2510.25015

\bibitem{AutoSpec24}
Wen, C., Cao, J., Su, J., Xu, Z., Qin, S., He, M., Li, H., Cheung, S., Tian,
  C.: Enchanting program specification synthesis by large language models using
  static analysis and program verification. In: Computer Aided Verification
  ({CAV} 2024). Lecture Notes in Computer Science, Springer (2024)

\bibitem{AutoVerus25}
Yang, C., Li, X., Misu, M.R.H., Yao, J., Cui, W., Gong, Y., Hawblitzel, C.,
  Lahiri, S.K., Lorch, J.R., Lu, S., Yang, F., Zhou, Z., Lu, S.: {AutoVerus}:
  Automated proof generation for {Rust} code. Proc. {ACM} Program. Lang.
  \textbf{9}({OOPSLA2}) (2025). \doi{10.1145/3763174}

\end{thebibliography}

\appendix
\Section{Per-Target Results}
\label{sec:appendix}

Tables~\ref{tab:sol} and~\ref{tab:terra} give, for each target and tactic, the mean cost per cell over the four replicates, mean input/output tokens in thousands (I/O k), the hybrid mean cost as a percentage of agent-only (vs. agent), and whether the specification killed every withheld mutant in all four replicates (\Checkmark) or a mutant survived (\XSolidBrush, subscripted with the count where more than one replicate failed); $\circ$ marks an unmeasured cell.
Input counts include cached tokens.
The targets above the middle rule have no loop that needs an invariant.

\begin{sidewaystable}
\centering
\caption{Sol~5.6 by target. Summed over the 26 targets, the tactics cost \$36.72, \$31.67, and \$31.81.}
\label{tab:sol}
\scriptsize
\setlength{\tabcolsep}{2.2pt}
\begin{tabular}{lrrcrrrcrrrc}
\toprule
& \multicolumn{3}{c}{agent-only} & \multicolumn{4}{c}{hybrid-guided} & \multicolumn{4}{c}{hybrid-flexible} \\
\cmidrule(lr){2-4} \cmidrule(lr){5-8} \cmidrule(lr){9-12}
target & cost & I/O k & ok & cost & vs. agent & I/O k & ok & cost & vs. agent & I/O k & ok \\
\midrule
|keep_alive| & \$0.517 & 588/3.7 & \Checkmark & \$0.148 & 29\% & 69/0.7 & \Checkmark & \$0.206 & 40\% & 135/1.2 & \Checkmark \\
|is_order_valid| & \$0.493 & 577/3.7 & \Checkmark & \$0.140 & 28\% & 73/0.8 & \Checkmark & \$0.209 & 42\% & 166/1.3 & \Checkmark \\
|available_transaction_queue_capacity| & \$0.239 & 212/2.4 & \Checkmark & \$0.131 & 55\% & 66/0.7 & \Checkmark & \$0.153 & 64\% & 120/1.1 & \Checkmark \\
|can_execute_with_timelock| & \$0.465 & 454/4.9 & \Checkmark & \$0.112 & 24\% & 67/0.7 & \Checkmark & \$0.163 & 35\% & 114/1.0 & \Checkmark \\
|is_taker_order| & \$0.218 & 188/1.7 & \Checkmark & \$0.118 & 54\% & 66/0.6 & \Checkmark & \$0.133 & 61\% & 92/0.9 & \Checkmark \\
|refill| & \$0.260 & 217/3.6 & \Checkmark & \$0.121 & 47\% & 67/0.7 & \Checkmark & \$0.178 & 69\% & 135/1.1 & \Checkmark \\
|request| & \$0.174 & 129/1.7 & \Checkmark & \$0.153 & 88\% & 87/1.4 & \Checkmark & \$0.210 & 121\% & 160/2.0 & \Checkmark \\
|validate_enough_stake| & \$0.291 & 240/4.5 & \Checkmark & \$0.228 & 78\% & 154/2.5 & \Checkmark & \$0.255 & 88\% & 227/2.6 & \Checkmark \\
|find_min_stake_required| & \$0.296 & 282/2.3 & \Checkmark & \$0.121 & 41\% & 77/0.9 & \Checkmark & \$0.208 & 70\% & 157/1.8 & \Checkmark \\
|update_config| & \$0.280 & 285/2.8 & \Checkmark & \$0.119 & 42\% & 71/0.7 & \Checkmark & \$0.172 & 61\% & 149/1.1 & \Checkmark \\
|match_order_and_get_next_from_bulk_order| & \$0.358 & 332/4.2 & \Checkmark & \$0.123 & 34\% & 69/0.7 & \Checkmark & \$0.191 & 53\% & 131/1.0 & \Checkmark \\
|reinsert_order_into_bulk_order| & \$0.328 & 310/2.8 & \Checkmark & \$0.125 & 38\% & 67/0.7 & \Checkmark & \$0.177 & 54\% & 129/1.1 & \Checkmark \\
|new_bulk_order_request_with_sanitization| & \$0.280 & 213/3.3 & \Checkmark & \$0.207 & 74\% & 110/2.3 & \Checkmark & \$0.255 & 91\% & 160/2.9 & \Checkmark \\
\midrule
|validate_scheme| & \$0.349 & 309/4.2 & \Checkmark & \$0.484 & 139\% & 486/6.7 & \Checkmark & \$0.366 & 105\% & 367/4.4 & \Checkmark \\
|upsert_provider_jwks| & \$0.565 & 574/7.5 & \Checkmark & \$0.314 & 56\% & 276/4.0 & \XSolidBrush$_{3}$ & \$0.468 & 83\% & 422/7.9 & $\circ_{2}$ \\
|get_pending_transactions| & \$0.705 & 779/6.6 & \Checkmark & \$0.739 & 105\% & 835/10.2 & \Checkmark & \$0.725 & 103\% & 833/8.7 & \Checkmark \\
|partition| & \$0.504 & 508/6.3 & \XSolidBrush & \$1.032 & 205\% & 1449/11.3 & \XSolidBrush & \$0.589 & 117\% & 658/7.2 & \XSolidBrush \\
|select| & \$0.374 & 273/5.6 & \Checkmark & \$0.472 & 126\% & 392/7.7 & $\circ$ & \$0.475 & 127\% & 385/7.0 & $\circ$ \\
|split_signature_bytes| & \$0.282 & 266/2.4 & \Checkmark & \$0.422 & 150\% & 465/4.3 & \Checkmark & \$0.378 & 134\% & 393/3.6 & \Checkmark \\
|new_tiers| & \$0.342 & 292/4.5 & \Checkmark & \$0.438 & 128\% & 388/6.5 & \Checkmark & \$0.414 & 121\% & 371/5.1 & \Checkmark \\
|validate_tiers| & \$0.389 & 418/3.8 & \Checkmark & \$0.416 & 107\% & 454/4.5 & \Checkmark & \$0.408 & 105\% & 472/4.1 & \Checkmark \\
|cancel_at_price_level| & \$0.342 & 266/4.0 & \Checkmark & \$0.303 & 88\% & 287/3.4 & \XSolidBrush & \$0.350 & 102\% & 264/3.6 & \XSolidBrush \\
|discard_price_crossing_levels| & \$0.272 & 235/2.7 & \Checkmark & \$0.262 & 96\% & 229/2.9 & \Checkmark & \$0.279 & 102\% & 261/3.1 & \Checkmark \\
|validate_not_zero_sizes| & \$0.240 & 197/1.5 & \Checkmark & \$0.224 & 93\% & 210/1.9 & \Checkmark & \$0.245 & 102\% & 261/2.1 & \Checkmark \\
|validate_price_ordering| & \$0.264 & 233/2.7 & \Checkmark & \$0.269 & 102\% & 253/3.1 & \Checkmark & \$0.309 & 117\% & 301/2.9 & \Checkmark \\
|range_with_step| & \$0.356 & 290/4.9 & \Checkmark & \$0.697 & 196\% & 711/10.2 & \Checkmark & \$0.438 & 123\% & 408/6.7 & \Checkmark \\
\midrule
\textbf{all 26} & \$0.353 & 333/3.8 & 103/104 & \$0.304 & 86\% & 288/3.5 & 98/104 & \$0.306 & 87\% & 280/3.3 & 98/104 \\
\bottomrule
\end{tabular}

\end{sidewaystable}

\begin{sidewaystable}
\centering
\caption{Terra~5.6 by target. Summed over the 26 targets, the tactics cost \$21.05, \$16.71, and \$18.79.}
\label{tab:terra}
\scriptsize
\setlength{\tabcolsep}{2.2pt}
\begin{tabular}{lrrcrrrcrrrc}
\toprule
& \multicolumn{3}{c}{agent-only} & \multicolumn{4}{c}{hybrid-guided} & \multicolumn{4}{c}{hybrid-flexible} \\
\cmidrule(lr){2-4} \cmidrule(lr){5-8} \cmidrule(lr){9-12}
target & cost & I/O k & ok & cost & vs. agent & I/O k & ok & cost & vs. agent & I/O k & ok \\
\midrule
|keep_alive| & \$0.273 & 585/4.2 & \Checkmark & \$0.060 & 22\% & 69/0.4 & \Checkmark & \$0.102 & 37\% & 176/0.9 & \Checkmark \\
|is_order_valid| & \$0.269 & 574/3.7 & \Checkmark & \$0.062 & 23\% & 67/0.4 & \Checkmark & \$0.110 & 41\% & 154/0.9 & \Checkmark \\
|available_transaction_queue_capacity| & \$0.113 & 182/1.9 & \Checkmark & \$0.059 & 52\% & 71/0.4 & \Checkmark & \$0.085 & 75\% & 124/0.7 & \Checkmark \\
|can_execute_with_timelock| & \$0.239 & 470/4.4 & \XSolidBrush & \$0.059 & 25\% & 66/0.4 & \Checkmark & \$0.098 & 41\% & 165/0.9 & \Checkmark \\
|is_taker_order| & \$0.136 & 227/1.9 & \Checkmark & \$0.061 & 45\% & 66/0.4 & \Checkmark & \$0.088 & 64\% & 117/0.7 & \Checkmark \\
|refill| & \$0.168 & 267/3.9 & \Checkmark & \$0.060 & 35\% & 67/0.4 & \Checkmark & \$0.096 & 57\% & 141/0.9 & \Checkmark \\
|request| & \$0.100 & 151/1.6 & \Checkmark & \$0.059 & 59\% & 73/0.5 & \Checkmark & \$0.122 & 122\% & 171/1.9 & \Checkmark \\
|validate_enough_stake| & \$0.217 & 358/5.1 & \Checkmark & \$0.141 & 65\% & 239/2.4 & \Checkmark & \$0.121 & 56\% & 223/1.9 & \Checkmark \\
|find_min_stake_required| & \$0.175 & 321/2.3 & \Checkmark & \$0.083 & 47\% & 114/1.1 & \Checkmark & \$0.118 & 68\% & 201/1.7 & \Checkmark \\
|update_config| & \$0.159 & 289/2.7 & \Checkmark & \$0.065 & 41\% & 69/0.4 & \Checkmark & \$0.105 & 66\% & 175/0.9 & \Checkmark \\
|match_order_and_get_next_from_bulk_order| & \$0.257 & 517/4.8 & \Checkmark & \$0.065 & 25\% & 68/0.4 & \Checkmark & \$0.086 & 34\% & 127/0.8 & \Checkmark \\
|reinsert_order_into_bulk_order| & \$0.192 & 340/3.2 & \Checkmark & \$0.057 & 30\% & 66/0.4 & \Checkmark & \$0.079 & 41\% & 118/0.7 & \Checkmark \\
|new_bulk_order_request_with_sanitization| & \$0.173 & 263/3.8 & \Checkmark & \$0.059 & 34\% & 75/0.5 & \Checkmark & \$0.095 & 55\% & 141/0.8 & \Checkmark \\
\midrule
|validate_scheme| & \$0.142 & 224/3.0 & \Checkmark & \$0.173 & 122\% & 292/3.5 & \Checkmark & \$0.202 & 142\% & 361/4.3 & \Checkmark \\
|upsert_provider_jwks| & \$0.310 & 543/8.1 & $\circ$ & \$0.220 & 71\% & 346/5.1 & \XSolidBrush$_{2}$ & \$0.335 & 108\% & 603/9.1 & \XSolidBrush \\
|get_pending_transactions| & \$0.306 & 663/5.1 & \Checkmark & \$0.325 & 106\% & 644/7.8 & \Checkmark & \$0.392 & 128\% & 850/9.2 & \Checkmark \\
|partition| & \$0.198 & 372/3.7 & \XSolidBrush$_{4}$ & \$0.476 & 240\% & 1129/11.3 & \XSolidBrush$_{2}$ & \$0.255 & 129\% & 489/5.8 & \XSolidBrush$_{4}$ \\
|select| & \$0.338 & 612/9.0 & $\circ$ & \$0.558 & 165\% & 1224/15.0 & $\circ$ & \$0.435 & 129\% & 906/10.3 & \Checkmark \\
|split_signature_bytes| & \$0.185 & 351/2.8 & \Checkmark & \$0.172 & 93\% & 339/2.6 & \Checkmark & \$0.193 & 105\% & 355/3.7 & \Checkmark \\
|new_tiers| & \$0.232 & 412/5.9 & \Checkmark & \$0.238 & 102\% & 394/5.5 & \Checkmark & \$0.320 & 138\% & 660/7.4 & \Checkmark \\
|validate_tiers| & \$0.195 & 381/3.4 & \Checkmark & \$0.260 & 133\% & 595/4.8 & \Checkmark & \$0.320 & 164\% & 745/5.7 & \Checkmark \\
|cancel_at_price_level| & \$0.192 & 314/4.3 & \Checkmark & \$0.194 & 101\% & 344/4.3 & \XSolidBrush$_{2}$ & \$0.188 & 98\% & 326/3.6 & \XSolidBrush \\
|discard_price_crossing_levels| & \$0.194 & 343/3.5 & \Checkmark & \$0.144 & 74\% & 220/2.5 & \Checkmark & \$0.184 & 95\% & 317/3.3 & \Checkmark \\
|validate_not_zero_sizes| & \$0.125 & 202/1.6 & \Checkmark & \$0.144 & 115\% & 225/1.7 & \Checkmark & \$0.140 & 112\% & 272/2.1 & \Checkmark \\
|validate_price_ordering| & \$0.159 & 255/2.2 & \Checkmark & \$0.144 & 91\% & 224/2.3 & \Checkmark & \$0.164 & 103\% & 293/2.8 & \Checkmark \\
|range_with_step| & \$0.216 & 331/5.4 & \Checkmark & \$0.242 & 112\% & 426/5.8 & \Checkmark & \$0.264 & 122\% & 489/6.7 & \Checkmark \\
\midrule
\textbf{all 26} & \$0.202 & 367/3.9 & 97/104 & \$0.161 & 79\% & 289/3.1 & 97/104 & \$0.181 & 89\% & 335/3.4 & 98/104 \\
\bottomrule
\end{tabular}

\end{sidewaystable}

\end{document}